\documentclass{nature}

\usepackage{graphicx}
\makeatletter
\let\saved@includegraphics\includegraphics
\AtBeginDocument{\let\includegraphics\saved@includegraphics}
\renewenvironment*{figure}{\@float{figure}}{\end@float}
\makeatother

\usepackage{braket}
\usepackage{amsmath}

\title{Optical clock intercomparison with $6\times 10^{-19}$ precision in one hour.}

\author{E. Oelker$^{1}$, R. B. Hutson$^1$, C. J. Kennedy$^1$, L. Sonderhouse$^1$, T. Bothwell$^1$, A. Goban$^1$, D. Kedar$^1$, C. Sanner$^1$, J. M. Robinson$^1$, G. E. Marti$^1$, D. G. Matei$^{2,}$\footnote{Present address: Horia Hulubei National Institute of Physics and Nuclear Engineering, Reactorului 30, 077125 Magurele, Romania}, T. Legero$^2$, M. Giunta$^{3,4}$, R. Holzwarth$^{3,4}$, F. Riehle$^2$, U. Sterr$^2$, \& J. Ye$^1$}

\begin{document}

\maketitle

\begin{affiliations}
 \item JILA, National Institute of Standards and Technology and University of Colorado, Department of Physics, University of Colorado, Boulder, CO 80309, USA.
 \item Physikalisch-Technische Bundesanstalt, Bundesallee 100, 38116 Braunschweig, Germany.
 \item Menlo Systems GmbH, Am Klopferspitz 19a, 82152 Martinsried, Germany.
 \item Max-Planck-Institut f{\"u}r Quantenoptik, Garching, Germany.
\end{affiliations}

\begin{abstract}
Improvements in atom-light coherence are foundational to progress in quantum information science\cite{Wineland2003}, quantum optics\cite{Norcia2018}, and precision metrology\cite{Bloom2014}.  Optical atomic clocks require local oscillators with exceptional optical coherence due to the challenge of performing spectroscopy on their ultra-narrow linewidth clock transitions. Advances in laser stabilization have thus enabled rapid progress in clock precision\cite{Nicholson2012,Schioppo2017}.  A new class of ultrastable lasers based on cryogenic silicon reference cavities\cite{Kessler2012} has recently demonstrated the longest optical coherence times to date\cite{Matei2017}.  In this work we utilize such a local oscillator, along with a state-of-the-art frequency comb for coherence transfer, with two Sr optical lattice clocks to achieve an unprecedented level of clock stability. Through an anti-synchronous comparison, the fractional instability of both clocks is assessed to be $4.8\times 10^{-17}/\sqrt{\tau}$ for an averaging time $\tau$ in seconds. Synchronous interrogation reveals a quantum projection noise dominated instability of $3.5(2)\times10^{-17}/\sqrt{\tau}$, resulting in a precision of $5.8(3)\times 10^{-19}$ after a single hour of averaging.  The ability to measure sub-$10^{-18}$ level frequency shifts in such short timescales will impact a wide range of applications for clocks in quantum sensing\cite{Chou2010_2,Takano2016,Grotti2018,Kolkowitz2016} and fundamental physics\cite{Huntemann2014, Godun2014, Sanner2018,Derevianko2014,Dilaton2015}. For example, this precision allows one to resolve the gravitational red shift from a 1 cm elevation change in only 20 minutes.
\end{abstract}

Atomic frequency standards play a vital role in a growing number of technological and scientific endeavours.  Atomic clocks based on microwave transitions currently define the SI (international system of units) second and are used extensively for network synchronization and satellite based navigation systems such as the global positioning system (GPS).  Over the last two decades, a new generation of atomic clocks based on narrow linewidth optical transitions has reached maturity and now surpasses its microwave predecessors in both accuracy and precision \cite{Bloom2014,Nicholson2015,Huntemann2016,Takano2016,McGrew2018}.  Further improvements in clock performance help pave the way for the eventual redefinition of the SI second in terms of an optical transition.  Optical clocks are also ideal for observing physical phenomena that cause minute alterations in the clock transition frequency.  They can be used to detect possible drift in the values of fundamental constants\cite{Huntemann2014,Godun2014}, enable relativistic geodesy\cite{Chou2010_2,Takano2016,Grotti2018}, test fundamental symmetries\cite{Sanner2018}, and even potentially detect passing gravitational waves\cite{Kolkowitz2016} or dark matter\cite{Derevianko2014,Dilaton2015}. 

All of these exciting applications have one commonality: they require ultrastable frequency comparison measurements between two or more atomic clocks. The stability is a measure of the noise of a frequency standard and is often expressed as fractional frequency fluctuations $\delta \nu_{c}/\nu_{c}$ where $\nu_{c}$ is the clock transition frequency. Quantum projection noise (QPN) determines the standard quantum limit (SQL) for the clock instability\cite{QPN}.  For Rabi spectroscopy, this limit is given by 

\begin{equation}\label{eq:QPN}
\sigma_{QPN}(\tau) = \frac{0.264}{\nu_{\textrm{c}} T_p}\sqrt{\frac{T_c}{N\tau}}
\end{equation}

\noindent where $T_{p}$ is the Rabi spectroscopy time, $T_{c}$ is the clock cycle time, $N$ is the number of atoms.

In neutral atom optical lattice clocks thousands of atoms are interrogated simultaneously leading to a favorably small SQL for clock instability.  However, optical lattice clocks generally operate with an instability above this limit due to noise on the laser used to probe the clock transition.  The clock transition typically cannot be measured continuously since time is required for sample preparation and state detection.  During the interval between successive measurements, hereafter referred to as \textit{dead time}, the local oscillator serves as a flywheel and largely determines short-term stability of the clock.  Due to the intermittency of the measurements, high frequency noise from the probe laser is aliased down to lower frequencies degrading the clock stability at long averaging times.  To date this technical noise coupling, known as the Dick effect, has limited single-ensemble instability\cite{Nicholson2012,Bloom2014,Schioppo2017,McGrew2018,Masoudi2015} to the low $10^{-16}/\sqrt{\tau}$ level.  Advances in the stability of local oscillators improve optical clock performance by reducing the Dick effect and enabling longer interrogation times which lead to enhanced spectroscopic resolution\cite{Campbell2017} and a lower QPN bound. Progress in laser stabilization has also facilitated studies of spin-orbit coupling\cite{Kolkowitz2017} and many-body physics\cite{Martin2013,Xibo2014}.  

Recently, conventional optical local oscillators referenced to ultra-low expansion (ULE) glass cavities have reached the performance bound set by fundamental thermo-mechanical noise from their constituent materials\cite{Bishof2013,Hafner2015}.  To reach a new stability regime, we have developed an improved class of ultrastable lasers based on cryogenic silicon reference cavities\cite{Kessler2012,Matei2017,Wei2017,Robinson2018}.  The local oscillator used in this work is referenced to a 21 cm Si cavity operating at 124 K.  Its world record performance was evaluated by comparing two identical systems\cite{Matei2017}, each exhibiting a sub-10 mHz linewidth, a laser coherence time of 55 s for Rabi spectroscopy, and a thermal noise limited instability of $4\times 10^{-17}$.  


The experimental setup is depicted in Fig 1.  The local oscillator operates at 1542 nm because Si is not transparent at the Sr clock transition wavelength (698 nm).  An optical frequency comb is utilized to transfer the stability of the ultrastable laser at 1542 nm to 698 nm (see Methods).  A prestabilized laser at 698 nm is phase locked to the comb to complete the stability transfer process.  Local oscillator light is then delivered to both a one-dimensional (1D) and three-dimensional (3D) $^{87}\textrm{Sr}$ optical lattice clock via phase stabilized optical links.  For both clocks, an ensemble of $^{87}\textrm{Sr}$ atoms are first cooled with a magneto-optical trap (MOT) on the 32 MHz wide $^{1}S_{0} \rightarrow$  $^{1}P_{1}$ transition at 461 nm.  A second narrow-line MOT stage utilizing the 7.6 kHz wide $^{1}S_{0} \rightarrow$ $^{3}P_{1}$ intercombination transition at 689 nm cools the atoms to a few $\mu$K.

For the 1D clock, the atoms are then loaded into a 1D magic wavelength (813 nm) optical lattice with a depth of 180 $\textrm{E}_{\textrm{r}}$, where  $\textrm{E}_{\textrm{r}}$ is the lattice photon recoil energy, and spin polarized by optically pumping into one of the $\ket{^{1}S_{0}, m_{F} = \pm \frac{9}{2}}$ stretched states.  The lattice depth is ramped down to 32 E$_{\textrm{r}}$ resulting in atom numbers of 1500-2000 and sample temperatures below  1 $\mu$K.  A 500 mG (1 G = $10^{-4}$ Tesla) bias field is applied to resolve the hyperfine sublevels and Rabi spectroscopy is first performed on the $\ket{^{1}S_{0}, -\frac{9}{2}} \rightarrow \ket{^{3}P_{0}, -\frac{9}{2}}$ transition.  The excited state population fraction is measured on one side of resonance and then the clock laser frequency is stepped across resonance by one full-width half-maximum (FWHM) and the population fraction is measured again.  The difference between these two excitation fractions yields an error signal proportional to the offset of the clock laser frequency from resonance.   This offset is nulled using a digital servo to adjust the laser frequency to remain on resonance using an acousto-optic modulator (AOM).  This operation is interleaved with an identical locking scheme for the $\ket{^{1}S_{0}, +\frac{9}{2}} \rightarrow \ket{^{3}P_{0}, +\frac{9}{2}}$ transition.  Because the two transitions have opposing first-order Zeeman shifts, the correction signals for the two servos are averaged for all clock stability computations to reject frequency shifts due to fluctuations in the background magnetic field.  This technique is effective at rejecting slowly varying field noise but provides almost no rejection of fluctuations at Fourier frequencies above 50 mHz (see Supplemental Materials).  High frequency magnetic field fluctuations in the 1D system are typically below the few 100 $\mu$G level and only modestly impact the clock stability at short averaging times.

For the 3D clock, the atoms are transferred from the 689 nm MOT into a co-located crossed optical dipole trap at 1064 nm.  A trap depth of 1 $\mu$K is chosen to remove atoms above the recoil temperature (200 nK).  To minimize dead time in this work we elect to forego the evaporative cooling stage used previously\cite{Campbell2017}.  Approximately 20000 atoms are then loaded into a 3D magic wavelength lattice with a depth of 60 E$_{r}$ in each direction.  To further reduce short-term instability in the 3D system, clock spectroscopy is performed on the $\ket{^{1}S_{0}, \pm \frac{5}{2}} \rightarrow \ket{^{3}P_{0}, \pm \frac{3}{2}}$ transitions which are 22 times less sensitive to magnetic field noise\cite{Boyd2007} as depicted in Fig 3.  A 1 G bias field is applied followed by a $\pi$ pulse on the $\ket{^{1}S_{0}, \pm \frac{5}{2}} \rightarrow \ket{^{3}P_{0}, \pm \frac{3}{2}}$ transition. A 5 ms pulse of 461 nm light removes all remaining ground state atoms yielding a sample in the $\ket{^{3}P_{0}, \pm \frac{3}{2}}$ excited state with 1000 ($\textrm{m}_{F}$ = $-\frac{3}{2}$) or 2000 ($\textrm{m}_{F} = +\frac{3}{2}$) atoms.  Rabi spectroscopy is then performed from the excited state and the clock laser is stabilized to both the $\ket{^{1}S_{0}, \pm \frac{5}{2}} \rightarrow \ket{^{3}P_{0}, \pm \frac{3}{2}}$ transitions using a four-point locking sequence similar to the 1D system.

The frequency record for both clocks is logged and used to determine the single-clock instability by computing an Allan deviation of the relative fractional frequency fluctuations $\frac{1}{\sqrt{2}}(\nu_{\textrm{1D}}(t) - \nu_{\textrm{3D}}(t))/\nu_{c}$ (see Methods).

In order to compute an estimate of the Dick effect contribution to the instability of the clocks it is necessary to have a frequency-domain noise model of the local oscillator.  Fig. 2a depicts a cross-correlation measurement using hetrodyne beat signals between the clock laser and two reference ultrastable laser systems.  The reference systems are based on a 6 cm Si cavity operating at 4 K\cite{Wei2017,Robinson2018} and a room-temperature 40 cm ULE cavity\cite{Bishof2013} with thermal noise floors of $6.5\times 10^{-17}$ and $8.8\times 10^{-17}$ respectively.  From this measurement we compute the clock instability arising from the Dick effect\cite{Dick1988}

\begin{equation}
\sigma_{Dick}^2(\tau) = \frac{1}{\tau}\sum_{n=1}^{\infty} \biggl\lvert\frac{G(n/T_c)}{G(0)}\biggr\rvert^{2} S_y(n/T_c)
\end{equation}

\noindent where $S_y(f)$ is the power spectral density of fractional frequency fluctuations of the local oscillator (red trace in Fig. 2a) and $G(f)$ denotes the atomic sensitivity function at Fourier frequency $f$ (see Methods).  Fig. 3a shows the result of this computation as a function of Rabi interrogation time and dead time.  For the operating conditions used in Fig. 4a, we anticipate a Dick effect limit of $3.8\times 10^{-17}/\sqrt{\tau}$ and a QPN contribution of $2.9\times 10^{-17}/\sqrt{\tau}$ yielding a clock instability of $4.8\times 10^{-17}/\sqrt{\tau}$.

With such a low instability at 1 s, the clocks reach a precision where drifting systematic offsets that impact the clock performance can be easily identified after only a few hundred seconds of averaging. Active temperature stabilization is implemented on both systems to mitigate fluctuations in Stark shifts arising from blackbody radiation.  The 1D clock is operated with a shallow lattice depth and low atom number to achieve a density shift below $1\times10^{-17}$ ensuring that atom number fluctuations do not impact its stability.  This approach allows both clocks to average into the $10^{-19}$ decade without requiring frequency corrections to compensate for non-stationary systematics. 

The instability of the two systems is evaluated by performing both an anti-synchronous comparison where clock spectroscopy for the two systems has no temporal overlap and a synchronous comparison where clock spectroscopy is performed simultaneously.  An anti-synchronous comparison is sensitive to all sources of instability including the Dick effect, QPN, fluctuating systematics, and other sources of technical noise and provides an estimate of the performance both clocks achieve individually.  Both clocks operate with a spectroscopy time of 550 ms and a dead time of 570 ms, allowing for a buffer of 10 ms between the end of each clock pulse and the beginning of the next to ensure no overlap.  When operating in this configuration with a shared local oscillator, the added noise from the Dick effect in the two systems exhibits some anti-correlation and the measured instability is expected to be higher than the prediction from Eqn. 2 (see Methods).  As shown in Fig. 4a, this measurement attains an instability of $5.5(3)\times 10^{-17}/\sqrt{\tau}$, in excellent agreement with the anticipated limit of $5.7 \times 10^{-17}/\sqrt{\tau}$ from QPN and the Dick effect.  From this measurement and the noise model one can deduce a single-clock instability of $4.8\times 10^{-17}/\sqrt{\tau}$ for independent operation (i.e. when compared with an uncorrelated reference).  This is a threefold improvement over previously reported values\cite{Schioppo2017,Nicholson2012,McGrew2018,Bloom2014,Masoudi2015} and a nearly tenfold reduction in the corresponding averaging time.

Synchronous interrogation\cite{Katori2011,Nicholson2012,Katori2015,Takano2016,Schioppo2017,McGrew2018} can further improve stability through common mode rejection of the Dick effect.  In principle this can facilitate comparison measurements limited solely by QPN, though this requires disseminating the local oscillator to all atomic ensembles under interrogation with exceptionally high fidelity.  Synchronous interrogation has lead to stability improvements for clock comparisons within a single lab,\cite{Katori2011,Nicholson2012,Katori2015,Schioppo2017,McGrew2018} or over a regional fiber network\cite{Takano2016}, but is less practical over the longer distances envisaged for a future global network of optical clocks\cite{Riehle2017}.  Fig. 4b depicts a synchronous comparison where both clocks operate with 600 ms spectroscopy time and 570 ms dead time.  An unprecedented instability of $3.5(2)\times 10^{-17}/\sqrt{\tau}$ is observed, near the expected QPN limit of $2.8\times 10^{-17}/\sqrt{\tau}$. 

A third dataset, taken prior to optimizing the dead time of the 3D clock, is presented in Fig 4c.  This unsynchronized comparison has a different sensitivity to the Dick effect than the anti-synchronous comparison and provides an additional validation of the noise model (see Methods).  An instability of $6.4(1)\times 10^{-17}/\sqrt{\tau}$ results in an extrapolated precision of $5.3(1) \times 10^{-19}$ at the total measurement time of 14800 s.  The ability of both clocks to maintain stability near the model prediction over this extended duration suggests that drifting systematic offsets induce little, if any, additional instability.

The results presented here represent a significant advance in optical clock stability. An almost order-of-magnitude decrease in measurement time will accelerate the development of clocks with $10^{-19}$ level accuracy.  Future work will focus on improving the local oscillator both to reduce the Dick effect and to lower the QPN limit by extending the clock interrogation time.  Combining cryogenic silicon cavities with AlGaAs optical coatings\cite{Cole2013} provides a path towards laser stabilization at the low $10^{-18}$ level\cite{Matei2017,Wei2017} and laser coherence times approaching the natural lifetime of the clock excited state ($\sim$150 s).  Fully utilizing such a laser will require new techniques to improve the atomic coherence time\cite{Hutson2019} that, to date\cite{Marti2018}, has been limited to $<$10 s.  With further advances in laser and atomic coherence, clock stability can be improved by another order of magnitude or more. 

\begin{methods}

\subsection{Optical local oscillator}

A distributed feedback fiber laser operating at 1542 nm is stabilized to a 21 cm single-crystal silicon cavity with a finesse of 480,000 using the Pound-Drever-Hall (PDH) technique.  The cavity is actively temperature stabilized to 124 K where the coefficient of thermal expansion for silicon goes to zero.  The temperature actuation is realized with a custom low vibration cryostat using evaporated liquid nitrogen\cite{Kessler2012}.  The cavity is mounted vertically near the midplane using a three point support structure to minimize the impact of vibrations on its length stability.  The cavity chamber is maintained at a pressure of $10^{-10}$ mbar and located on an active vibration isolation platform.  Active residual amplitude modulation (RAM) control\cite{Wei2014} and intensity stabilization are implemented to optimize the long term stability.  By direct evaluation with a Sr clock and through extensive three-cornered hat measurements against two reference ultrastable lasers\cite{Wei2017,Bishof2013} the cavity instability was typically observed to be thermal noise limited at a level of $3.7\times 10^{-17}$ between 0.2-1000 seconds, though thermal noise performance occasionally extends out to longer averaging times (see  Fig. 2b and Fig. 5 in Supplemental Materials).  

A frequency-domain laser noise model based on a cross-correlation measurement is used to estimate the Dick effect limited stability for both clocks.  The cross-spectral density of the heterodyne beats between our local oscillator and two reference ultrastable lasers\cite{Wei2017,Bishof2013} is presented in Fig. 2a. Provided that the three systems are uncorrelated, the noise from the two reference systems will average away yielding an estimate of the power spectral density of the local oscillator's fractional frequency noise.  The resulting noise spectrum is used to construct the following model

\begin{equation}
S_{\textrm{laser}}(f) =  \frac{h_{-1}}{f} + h_0 + f^2 h_{2} + \sum_{i=1}^{N} \frac{a_i}{1+\left(\frac{f-f_i}{\Gamma_i/2}\right)^2}
\end{equation}

\noindent This model consists of a flicker frequency noise term ($h_{-1} = 1.5\times 10^{-33}  $), a white frequency noise term ($h_{0} = 4\times 10^{-34}$ $ \textrm{1/Hz}$), a white phase noise term ($h_{2} = 3 \times 10^{-36}$ $\textrm{1/Hz}^3 $), and a series of resonant features with parameters given in Table 1 in Supplemental Materials.

\subsection{Spectral purity transfer of the local oscillator}

The stability of the local oscillator is transferred from 1542 nm to the clock transition wavelength at 698 nm using an ultra-low noise optical frequency comb.  The comb is based on a femtosecond Er-doped polarization-maintaining fiber oscillator\cite{Hansel2017} operating at a repetition rate ($\textrm{f}_\textrm{rep}$) of 250 MHz and a center wavelength of 1560 nm.  The oscillator contains two intra-cavity electro-optic modulators that enable MHz bandwidth stabilization of both $\textrm{f}_\textrm{rep}$ and the carrier-envelope offset frequency ($\textrm{f}_\textrm{ceo}$) with minimal cross-talk\cite{Hansel2017_2}.  Additional low-bandwidth piezo and thermal actuators extend the tuning range of $\textrm{f}_\textrm{rep}$ enabling the comb to remain locked for months at a time.

The output of the femtosecond laser is amplified and split to seed two distinct comb branches. The first arm generates an octave-spanning spectrum used to measure $\textrm{f}_\textrm{ceo}$ via an f-2f interferometer. The second branch is designed to achieve low-noise spectral purity transfer by providing outputs at both 1542 nm and 698 nm with minimal differential path length fluctuations.  A narrow-band spectrum at 1397 nm is generated using four-wave mixing in a highly-nonlinear fiber.  A portion of the transmitted seed light is picked off to form a heterodyne beat with the local oscillator.  As depicted in Fig. 1, this signal is used to phase lock the comb to the local oscillator by actuating on $\textrm{f}_\textrm{rep}$.  The second output of this branch is frequency doubled using a PPLN crystal to produce a 2.5 nm FWHM spectrum centered at 698.4 nm.  A pre-stabilized external-cavity diode laser is then phase locked to the 698 nm comb output with a bandwidth of 1 kHz to complete the transfer process.  

The fidelity of the stability transfer is assessed by performing a multi-line comb comparison with an identical reference system in order to characterize the differential noise between the 1542 nm and 698 nm outputs\cite{Giunta2018}.  A well-defined comb line for each system is phase locked to a common optical reference at 1542.14 nm. The lock points for the $\textrm{f}_\textrm{ceo}$ and $\textrm{f}_\textrm{rep}$ stabilization servos are chosen such that the combs operate with the same repetition rate and that their spectral lines are offset from one another by 5 MHz across the near-infrared optical domain.  A heterodyne beat at 10 MHz is formed between the frequency doubled outputs of the two combs at 698 nm and counted using a dead-time free Lambda-type frequency counter with a 100 ms gate time.  Based on this qualification, both combs perform the transfer from 1542 nm to 698 nm with a fractional instability of $1.6 \times 10^{-18}$ at one second and any degradation in the stability of the clock laser is expected to be negligible (see Fig. 6 in Supplemental Materials).


\subsection{Atomic sensitivity function}

The Dick effect\cite{Dick1988} is a type of noise in clock spectroscopy arising from the aliasing of laser noise at integer multiples of the clock cycle frequency (n/T$_c$ where n is an integer) down to DC.  It sets a bound on the achievable fractional frequency instability of a single clock given by Eqn. 2.  $G(f)$ in Eqn. 2 is the Fourier transform of the atomic sensitivity function $g(t) = 2 \frac{\delta p_e}{\delta \phi}$, which describes the change induced in the excitation fraction ($\delta p_e$) by a phase shift in the local oscillator ($\delta \phi$) at time t.  For Rabi spectroscopy beginning at time $t=0$ the sensitivity function is given by\cite{Dick1988}

\begin{align}
\begin{split}
    g(t) &= \sin^2(\theta)\cos(\theta)\left[ \sin(\Omega_1(t))(1-\cos(\Omega_2(t))  + \sin(\Omega_2(t))(1-\cos(\Omega_1(t))) \right], \hspace{3mm}  0\leq t\leq T_{p} \\
    &= 0, \hspace{5mm} T_{p}<t\leq T_{c}
\end{split}
\end{align}

\noindent with $\theta = \pi/2-\arctan(2\Delta T_p)$ where $\Delta \approx \frac{0.4}{T_p}$ is the detuning from resonance for clock spectroscopy, $\Omega_1(t) = \pi\sqrt{1+(2\Delta T_p)}\frac{t}{T_p}$, and $\Omega_2(t) = \pi\sqrt{1+(2\Delta T_p)}\frac{T_p-t}{T_p}$.  

\subsection{Clock instability measurements}  To determine the instability of the 1D and 3D clocks, we measure the relative fractional frequency fluctuations between the systems.  Assuming that the two clocks have identical noise levels but are uncorrelated, the single clock instability can be inferred by analyzing

\begin{equation}
\frac{1}{\sqrt{2}}\frac{\nu_{\textrm{1D}}(t) - \nu_{\textrm{3D}}(t)}{\nu_{c}}
\end{equation}

\noindent where $\nu_{\textrm{1D}}(t)$ is the frequency record for the 1D clock, $\nu_{\textrm{3D}}(t)$ is the frequency record for the 3D clock, and $\nu_{c}$ is the Sr clock transition frequency.  For the unsynchronized comparison in Fig. 4c, the frequency record of the 1D clock is interpolated to match the measurement times of the 3D clock prior to computing the stability.

In similar comparison measurements\cite{Nicholson2012,Bloom2014,Schioppo2017,McGrew2018}, the frequency record of each clock is typically derived from the frequency corrections applied by the clock servo to keep the laser on resonance.  However the atomic servos have a finite bandwidth (30 mHz) and short term fluctuations between the laser frequency and the atomic resonance are not fully accounted for in the correction signal.  One can gain additional information about the noise present at higher Fourier frequencies by also examining the servo error signal.  In this work, the error signal is used to correct the frequency record for each servo as follows\cite{Campbell2017}

\begin{equation}
\nu(t_i) = \nu_\textrm{corr}(t_{i-1}) + \frac{\Gamma}{2A}\times \delta p_e(t_i)
\end{equation}

\noindent where $t_i$ is the measurement time for the $i^\textrm{th}$ clock cycle, $\nu_\textrm{corr}(t_{i-1})$ is the servo correction signal, $\Gamma$ is the FWHM linewidth, $A$ is the excitation fraction at the peak of the atomic resonance, and $\delta p_e(t_i)$ is the error signal given by the difference between the measured excitation fraction on the left and right side of resonance.  After correcting for the error signal one would expect an Allan deviation limited solely by QPN and the Dick effect to follow a constant $1/\sqrt{\tau}$ slope at all averaging times.  The fact that the traces in Fig. 4 deviate from this behavior at short averaging times indicates the presence of additional short term noise.  In all three comparisons, the noise on the 1D clock exceeds that of the 3D clock at averaging times below 40 s.  High frequency magnetic field induced noise is not efficiently rejected by the spectroscopy sequence (see Supplemental Materials) and is a plausible cause of excess instability at these averaging times in a clock with this level of precision.  Implementing clock spectroscopy on the $\ket{^{1}S_{0}, \pm \frac{5}{2}} \rightarrow  \ket{^{3}P_{0}, \pm \frac{3}{2}}$ transitions proved impractical in the 1D system due to an insufficient atom number after state preparation.

To give this short term noise sufficient time to average away, a weighted $1/\sqrt{\tau}$ fit is applied to Allan deviation points beyond 80 s when determining the instability in each comparison measurement. All instability calculations use the total Allan deviation with error bars computed according to Ref.\cite{Howe2000} 

\subsection{Dick effect prediction}  While the assumption that the noise contributions from the two clocks are uncorrelated is correct for the QPN contributions that dominate the synchronous comparison, it is only approximately true for the anti-synchronized and unsynchronized comparisons because the two clocks share the same local oscillator and correlations remain even when the clock interrogations have no temporal overlap.  To quantify this effect for the anti-synchronized comparison, one can derive an alternate form of Eqn. 2 to compute the Dick effect limit for a comparison between two clocks with sequences that are identical other than a timing offset of $\Delta t$.  Examining Eqns. 4 and 5 one notes that this scenario is equivalent to a single clock with an effective sensitivity function of $[g(t+\Delta t) - g(t)]/\sqrt{2}$. Using the shift theorem for the discrete Fourier transform, the n$^\textrm{th}$ Fourier transform coefficient of $g(t+\Delta t)$ is given by $G(f_{n})e^{2\pi in\Delta t/T_{c}} $. One can then modify Eqn. 2 as follows

\begin{align}
\begin{split}
\sigma^{2}_{\textrm{Dick}}(\tau) &= \frac{1}{2\tau}\sum_{n=1}^{\infty}|e^{2\pi in\Delta t/T_{c}} - 1|^2 \biggl\lvert\frac{G(n/T_c)}{G(0)}\biggr\rvert^{2} S_y(n/T_c) \\
&= \frac{2}{\tau}\sum_{n=1}^{\infty}  \sin^{2}\biggl(\frac{\pi n \Delta t}{T_{c}}\biggr)  \biggl\lvert\frac{G(n/T_c)}{G(0)}\biggr\rvert^{2} S_y(n/T_c)
\end{split}
\end{align}

\noindent For the anti-synchronous comparison ($\Delta t = 560$ ms) the laser noise model predicts a Dick effect limit of $5.0\times 10^{-17}/\sqrt{\tau}$, 30\% higher than the prediction for independent clock operation.  The $n=1$ term in Eqn. 7, which dominates the sum, is inflated by sampling noise at a Fourier frequency of $1/T_c$ using two clocks with $\Delta t = T_c/2$.  The noise at this Fourier frequency ends up being strongly anti-correlated between the two systems.  The anti-synchronous result therefore represents an upper limit for the single clock instability.  Adding this in quadrature with QPN yields a prediction of $5.7\times 10^{-17}/\sqrt{\tau}$, in excellent agreement with the measured instability of $5.5(3)\times 10^{-17}/\sqrt{\tau}$. 

For the unsynchronized comparison in Fig. 4c the 1D and 3D clocks operate with different cycle and Rabi interrogation times and Eqn. 2 does not apply.  To estimate the Dick effect a frequency record for both clocks is computed numerically using simulated laser noise data based on the noise model and the stability is calculated according to Eqn. 5.  This results in an anticipated stability of $6.0\times 10^{-17}/\sqrt{\tau}$, in reasonable agreement with the measured value.

\end{methods}


\begin{thebibliography}{1}
\bibitem{Wineland2003} Leibfried, D., Blatt, R., Monroe, C. \& Wineland, D.  Quantum dynamics of single trapped ions. \emph{Rev. Mod. Phys.} \textbf{75,} 281–324 (2003).

\bibitem{Norcia2018}  Norcia, M. A. \emph{et al.}  Frequency measurements of superradiance from the strontium clock transition. \emph{Phys. Rev. X} \textbf{8,}  021036 (2018).

\bibitem{Bloom2014} Bloom, B. J., Nicholson, T. L., Williams, J. R., Campbell, S. L., Bishof, M., Zhang, X., Zhang, W., Bromley, S. L., \& Ye, J. An optical lattice clock with accuracy and stability at the $10^{-18}$ level.  \emph{Nature} \textbf{506}, 71 (2014).

\bibitem{Schioppo2017} Schioppo, M. \emph{et al.}  Ultra-stable optical clock with two cold-atom ensembles. \emph{Nat. Photonics} \textbf{11,} 48-52 (2016).

\bibitem{Nicholson2012} Nicholson, T. L. \emph{et al.}  Comparison of two independent Sr optical clocks with $1\times10^{− 17}$ stability at $10^3$ s. \emph{Phys. Rev. Lett.} \textbf{109,} 230801 (2012).

\bibitem{Kessler2012} Kessler, T. \emph{et al.}  A sub-40-mHz-linewidth laser based on a silicon
single-crystal optical cavity. \emph{Nat. Photonics} \textbf{6,} 687–692 (2012).

\bibitem{Matei2017} Matei, D. G. \emph{et al.}  1.5 $\mu$m lasers with sub-10 mHz linewidth. \emph{Phys. Rev. Lett.} \textbf{118,} 263202 (2017).

\bibitem{Chou2010_2} Chou, C. W., Hume, D. B., Rosenband, T. \& Wineland, D. J. Optical Clocks and Relativity.  \emph{Science} \textbf{329,} 1630-1633 (2010).

\bibitem{Takano2016} Takano, T. \emph{et al.}  Geopotential measurements with synchronously linked optical lattice clocks. \emph{Nat. Photonics} \textbf{10,} 662–666 (2016).

\bibitem{Grotti2018} Grotti, J. \emph{et al.}  Geodesy and metrology with a transportable optical clock.  \emph{Nat. Phys.} \textbf{14,} 437-441 (2018).

\bibitem{Kolkowitz2016} Kolkowitz, S. \emph{et al.} Gravitational wave detection with optical lattice atomic clocks. \emph{Phys. Rev. D} \textbf{94}, 124043 (2016).

\bibitem{Huntemann2014}Huntemann, N. \emph{et al.}  Improved limit on a temporal variation of m$_p$/m$_e$ from comparisons of Yb$^+$ and Cs atomic clocks.  \emph{Phys. Rev. Lett} \textbf{113,} 210802 (2014).

\bibitem{Godun2014}Godun, R. M. \emph{et al.}  Frequency ratio of two optical clock transitions in $^{171}$Yb$^+$ and constraints
on the time variation of fundamental constants.  \emph{Phys. Rev. Lett} \textbf{113,} 210801 (2014).

\bibitem{Sanner2018} Sanner, C. \emph{et al.}
Optical clock comparison test of Lorentz symmetry. Preprint at https://arxiv.org/abs/1809.10742 (2018).

\bibitem{Derevianko2014} Derevianko, A. \& Pospelov, M. Hunting for topological dark matter with atomic clocks. \emph{Nat. Phys.} \textbf{10}, 933–936 (2014).

\bibitem{Dilaton2015}Arvanitaki, A., Huang, J. \& Van Tilburg, K. Searching for dilaton dark matter with atomic clocks.  \emph{Phys. Rev. D} \textbf{91}, 015015 (2015).

\bibitem{Nicholson2015} Nicholson, T. L. \emph{et al.}  Systematic evaluation of an atomic clock at $2\times10^{-18}$ total uncertainty. \emph{Nat. Commun.} \textbf{6}, 6896 (2015).

\bibitem{Huntemann2016} Huntemann, N., Sanner, C., Lipphardt, B., Tamm, C. \& Peik, E. Single-ion atomic clock with $3\times10^{-18}$ systematic uncertainty. \emph{Phys. Rev. Lett.} \textbf{116,} 063001 (2016).

\bibitem{McGrew2018} McGrew, W. F. \emph{et al.}  Atomic clock performance enabling geodesy below
the centimetre level. \emph{Nature} \textbf{564,} 87–90 (2018).

\bibitem{QPN} Itano, W. M. \emph{et al.} Quantum projection noise: population fluctuations in two-level systems. \emph{Phys. Rev. A} \textbf{47,} 3554–3570 (1993).

\bibitem{Masoudi2015} Al-Masoudi, A. \emph{et al.} Noise and instability of an optical lattice clock.  \emph{Phys. Rev. A} \textbf{92,} 063814 (2015).

\bibitem{Campbell2017} Campbell, S. L. \emph{et al.} A Fermi-degenerate three-dimensional optical lattice clock. \emph{Science}, \textbf{358,} 90–94 (2017).

\bibitem{Kolkowitz2017} Kolkowitz, S. \emph{et al.}
Spin-orbit-coupled fermions in an optical lattice clock.  \emph{Nature} \textbf{542,} 66–70 (2017).

 \bibitem{Xibo2014} Zhang, X \emph{et al.} Spectroscopic observation of SU(N)-symmetric interactions in Sr orbital magnetism.  \emph{Science} \textbf{345,} 1467-1473 (2014).

\bibitem{Martin2013} Martin, M. J. \emph{et al.}  A Quantum many-body spin system in an optical lattice clock. \emph{Science} \textbf{341,} 632-636 (2013).

\bibitem{Bishof2013} Bishof, M., Zhang, X., Martin, M. J. \& Ye, J.  Optical Spectrum Analyzer with Quantum-Limited Noise Floor.  \emph{Phys. Rev. Lett.} \textbf{111,} 093604 (2013).

\bibitem{Hafner2015} H{\"a}fner, S. \emph{et al.} $8\times10^{−17}$ fractional laser frequency instability with a long room-temperature cavity. \emph{Opt. Lett.} \textbf{40,} 2112-2115 (2015).

\bibitem{Wei2017} Zhang, W. \emph{et al.}  Ultrastable silicon cavity in a continuously operating closed-cycle cryostat at 4K. \emph{Phys. Rev. Lett.} \textbf{119,} 243601 (2017).

\bibitem{Robinson2018} Robinson, J. M. \emph{et al.}  Crystaline optical cavity at 4K with thermal noise limited instability and ultralow drift.  \emph{Optica} in press (2019).

\bibitem{Boyd2007} Boyd, M. M. \emph{et al.}  Nuclear spin effects in optical lattice clocks.  \emph{Phys. Rev. A.} \textbf{76,} 022510 (2007).

\bibitem{Dick1988} Dick, J. G.  Local oscillator induced instabilities in trapped ion frequency standards. in  \emph{Proceedings of 19th Annual Precise Time
and Time Interval Meeting}, pp. 133-147 (US Naval Observatory, 1988) at https://tycho.usno.navy.mil/ptti/ptti1987.html.

\bibitem{Katori2011} Takamoto, M., Takano, T. \& Katori, H. Frequency comparison of optical lattice clocks beyond the Dick limit. \emph{Nat. Photonics} \textbf{5,} 288–292 (2011).

\bibitem{Katori2015} Ushijima, I., Takamoto, M., Das, M., Ohkubo, T. \& Katori, H. Cryogenic optical lattice clocks. \emph{Nat. Photonics} \textbf{9,} 185–189 (2015).

\bibitem{Riehle2017} Riehle, F. Optical clock networks. \emph{Nat. Photonics} \textbf{11,} 25–31 (2017).


\bibitem{Cole2013} Cole, G. D. \emph{et al.}  Tenfold reduction of Brownian noise in high-reflectivity optical coatings. \emph{Nat. Photonics} \textbf{7,} 644 (2013). 

\bibitem{Hutson2019} Hutson, R. B. \emph{et al.}  Engineering quantum states of matter for atomic clocks in shallow optical lattices.  In preparation (2019).

\bibitem{Marti2018} Marti, G. E. \emph{et al.}  Imaging optical frequencies with 100 $\mu$Hz precision and 1.1 $\mu$m resolution.  \emph{Phys. Rev. Lett.}  \textbf{120,} 103201 (2018).


\bibitem{Wei2014} Zhang, W. \emph{et al.}  Reduction of residual amplitude modulation to $1\times 10^{-6}$ for frequency modulation and laser stabilization.  \emph{Opt. Lett.} \textbf{39} 1980 (2014).

\bibitem{Hansel2017} H{\"a}nsel, W. \emph{et al.}  All polarization-maintaining fiber laser architecture for robust femtosecond pulse generation. \emph{Appl. Phys. B.} \textbf{123,} 41 (2017).

\bibitem{Hansel2017_2} H{\"a}nsel, W. \emph{et al.}  Rapid electro-optic control of the carrier-envelope-offset frequency for ultra-low noise frequency combs. in \emph{Joint Conference of the European Frequency and Time Forum and IEEE International Frequency Control Symposium} pp. 128-129 (2017).

\bibitem{Giunta2018} Giunta, M. \emph{et al.} Sub-mHz spectral purity transfer for next generation strontium optical atomic clocks, in \emph{Conference on Lasers and Electro-Optics.} SM1L.5 (2018).

\bibitem{Howe2000} Howe, D.  The total deviation approach to long-term characterization of frequency stability.  \emph{IEEE Trans. Ultrason., Ferroelectr. Freq. Control} \textbf{47,} 1102 (2000).
\end{thebibliography}


\begin{addendum}
 \item This work is supported by NIST, the Defense Advanced Research Projects Agency, the Air Force Office of Scientific Research Multidisciplinary University Research
Initiative, the NSF JILA Physics Frontier Center (NSF PHY-1734006), the Centre for Quantum Engineering and Space-Time Research (QUEST), and Physikalisch-Technische Bundesanstalt (PTB).  E.O. and C.K. are
supported by a postdoctoral fellowship from the National Research
Council, L.S. is supported by a National Defense Science and Engineering Graduate Fellowship, A.G. is supported by a fellowship from the Japan Society for the Promotion of Science, C. S. is supported by a fellowship from the Humboldt Foundation. T. L., D. G. M. and U. S. acknowledge support from the Quantum sensors (Q-SENSE) project, supported by the European Commission’s H2020 MSCA RISE
under Grant Agreement Number 69115.  M.G. and R.H. acknowledge support from the EU 
FP7 initial training network FACT (Future Atomic Clock Technology) and the DARPA Program in Ultrafast Laser Science and Engineering (P$\mu$reComb project) under contract no. W31P4Q-14-C-0050.  We thank J. Munez and J. Sherman for careful reading of this manuscript.
\item[Author Contributions]  E.O., R.B.H., C.K., T.B. L.S., C.S., D.K., A.G., J.M.R., G.E.M. and J.Y. contributed to the clock instability measurements.  E.O., J.M.R., L.S., C.K., T.B., D.K., D.G.M, T.L., F.R., U.S. and J.Y. worked on the Si cavity. L.S. and E.O. commissioned the laser stability transfer setup based on the Er frequency comb developed by M.G. and R.H. All authors contributed to scientific discussions and the writing of this manuscript.
 \item[Competing Interests] The authors declare that they have no
competing financial interests.
 \item[Correspondence] Correspondence and requests for materials should be addressed to 
 E. Oelker 

 (ericoelker@gmail.com) or J. Ye (Ye@jila.colorado.edu).
\end{addendum}

\begin{figure}\label{fig:apparatus}
\centering
\includegraphics[angle=270, width=1.0\textwidth]{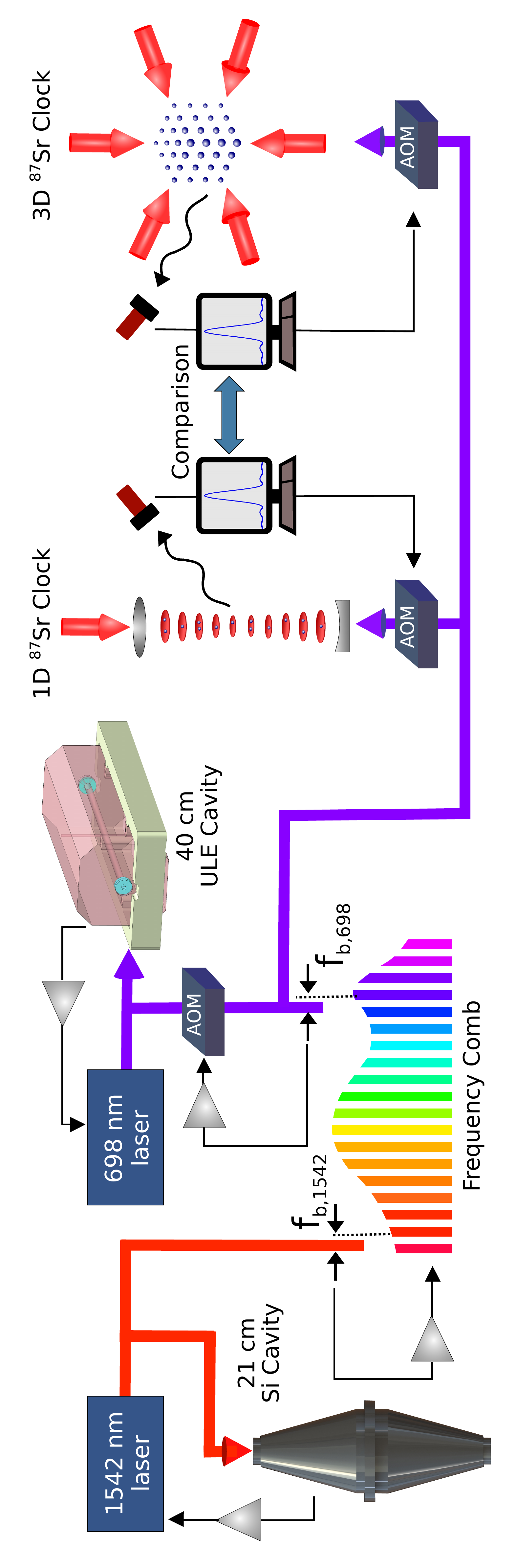}
\caption{\small \textbf{Experimental layout.}  A 1542 nm laser is stabilized to a 21 cm cryogenic Silicon resonator at 124 K using the Pound-Drever-Hall (PDH) technique.  The stability is transferred to a self-referenced femtosecond Er fiber frequency comb that is phase locked to the 1542 nm light by actuating on the comb repetition rate.  Light from the comb at 1396 nm is frequency doubled to the clock transition frequency at 698 nm and used to phase lock a prestabilized 698 nm laser to complete the stability transfer process.  698 nm light is then sent to both the 1D and 3D optical lattice clocks over optical links with active path length stabilization. Rabi spectroscopy of the clock transition is performed and frequency offsets between the laser and atomic resonance are detected by measuring the ensemble excited state fraction via atomic fluorescence using a photomultiplier tube. Two independent digital servos steer the laser frequency using AOMs to lock both probe paths to each atomic transition.  The frequency records for the two clocks are analyzed to perform intercomparison measurements.}
\end{figure}

\begin{figure}\label{fig:laser}
\centering
\includegraphics[width=0.9\textwidth]{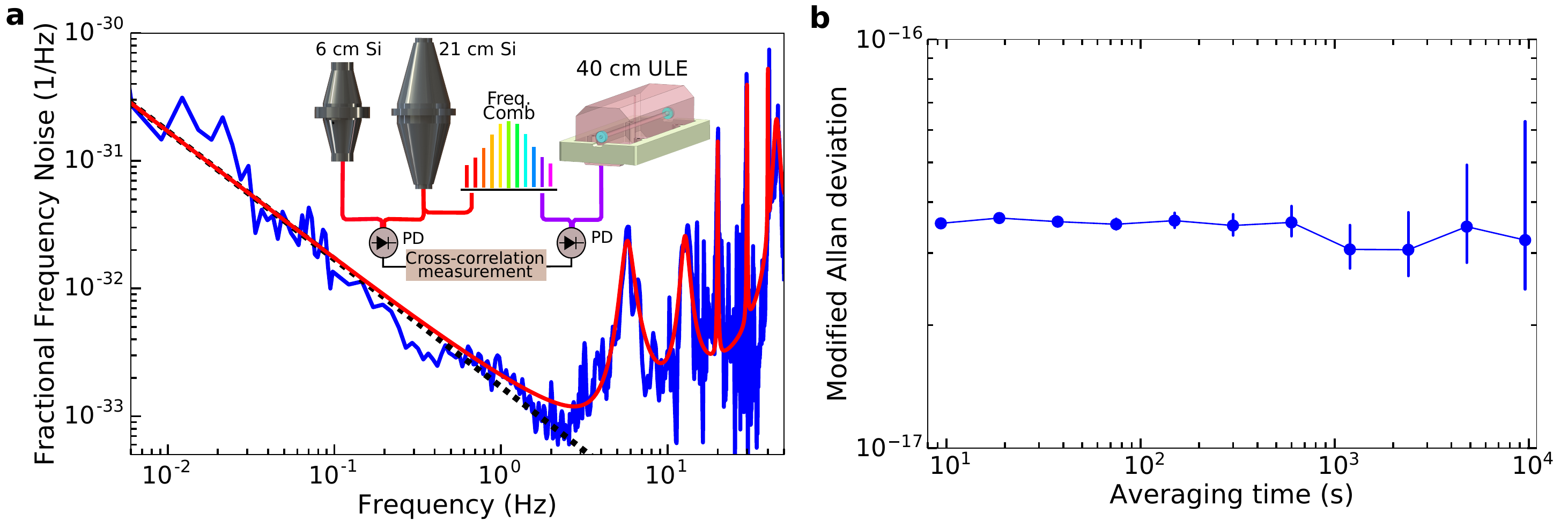}
\caption{\small \textbf{Laser stability characterization.  a)} The short-term stability of the 21 cm Si cavity is evaluated by measuring against two reference ultrastable lasers and computing a cross correlation.  The blue trace shows the cross-spectral density (CSD) of the two signals after 10 hours of averaging.  Provided that the three systems are uncorrelated, the noise from the reference systems will average away in the CSD yielding a frequency noise spectrum for the clock laser.  The resulting spectrum is consistent with the thermal noise floor (dashed black line) below 3 Hz.  A noise model based on this measurement (red trace) is used to predict the Dick effect limited stability of both clocks.  \textbf{b)}  Allan deviation of the frequency corrections applied by the atomic servo to stabilize the laser to the clock transition of the 1D lattice clock.  This constitutes a direct measurement of the clock laser stability at averaging times where the atomic servo is engaged ($>$ 30 s).  This 11 hour measurement showcases the improved long-term stability of the local oscillator which now maintains thermal noise limited performance at averaging times exceeding 1000 s.  A linear drift rate of 48 $\mu$Hz/s at 1542 nm was subtracted from the frequency record prior to computing the stability.}
\end{figure}

\begin{figure}\label{fig:dickEffectFields}
\centering
\includegraphics[width=1.0\textwidth]{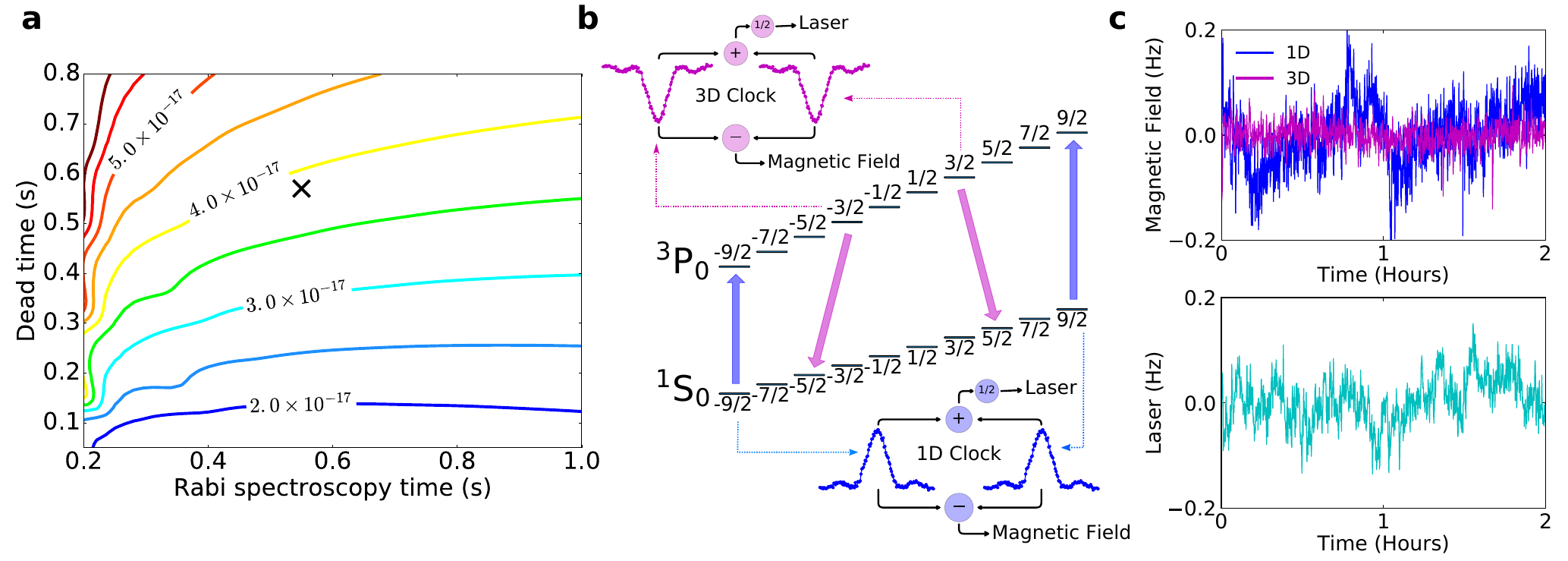}
\caption{\small \textbf{Impact of laser and magnetic field noise on clock instability.  a)}  Calculation of the Dick effect limited instability of an optical clock using Rabi spectroscopy assuming the laser noise model presented in Fig. 2a.  The contours show the instability at 1s as a function of Rabi spectroscopy time and dead time and the marker indicates the operating conditions used for the comparison measurement in Fig 4a.  \textbf{b)}  Clock spectroscopy for magnetic field noise cancellation.  Both systems reject magnetic field noise by alternately interrogating clock transitions in the hyperfine manifold with opposing Zeeman shifts and averaging the measured frequencies prior to computing the clock stability.  This technique is effective for cancelling noise from slow drift in the background field but does little to reject rapid fluctuations.  If these uncancelled fluctuations are comparable to laser noise they can degrade the clock stability.  The 1D clock utilizes $\pi$ transitions between the streched states $\ket{^{1}S_{0},m_F=\pm \frac{9}{2}} \rightarrow \ket{^{3}P_{0},m_F=\pm \frac{9}{2}}$ with a field sensitivity of 975 Hz/gauss.  To reject noise from field transients the 3D clock interrogates the $\ket{^{1}S_{0},m_F=\pm \frac{5}{2}} \rightarrow \ket{^{3}P_{0},m_F=\pm \frac{3}{2}}$ $\sigma_{\mp}$ transitions which are split by only 44 Hz/gauss.  \textbf{c)} Fluctuations in the first-order Zeeman shift induced by magnetic field noise in both clocks.  By probing the least magnetically sensitive transitions the 3D clock is largely immune to field noise.  A plot of local oscillator frequency fluctuations measured by the 1D clock after subtracting a linear frequency drift is included for comparison.}
\end{figure}

\begin{figure}\label{fig:stability}
\centering
\includegraphics[width=1.0\textwidth]{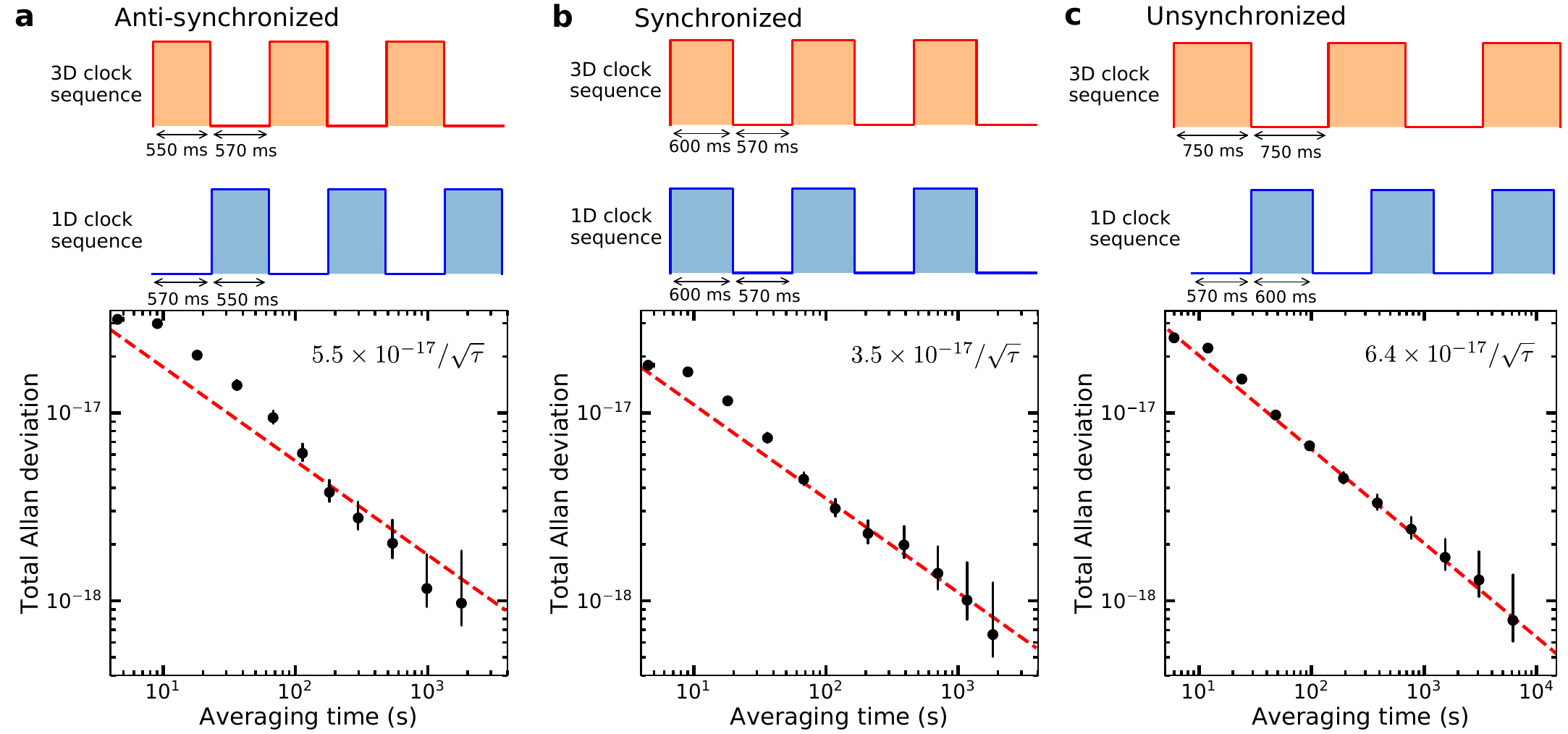}
\caption{\small \textbf{Intercomparison measurements between 1D and 3D clocks.  a)}  Anti-synchronous comparison.  The two clocks are run with identical cycle times but with the clock pulses carefully anti-synchronized to prevent common mode rejection of Dick effect.  The observed stability of $5.5(3)\times 10^{-17}/\sqrt{\tau}$ along with a detailed noise model allow one to infer a stability of $4.8\times 10^{-17}/\sqrt{\tau}$ for independent clock operation.  \textbf{b)}  Synchronous comparison.  The two clock pulses are carefully synchronized to provide maximal common mode rejection of the Dick effect.  A stability of $3.5(2)\times 10^{-17}/\sqrt{\tau}$ limited primarily by QPN is observed, resulting in a precision of $5.8(3)\times 10^{-19}$ after a single hour of averaging.  \textbf{c)}  Unsynchronized comparison.  This measurement demonstrates the capability of both clocks to average down with state of the art stability for a comparison spanning several hours.  By extrapolating a stability of $6.4(1)\times 10^{-17}/\sqrt{\tau}$ to the total measurement time of 14800s our clocks reach a precision of $5.3(1)\times10^{-19}$.  All error bars represent $1\sigma$ confidence intervals.}
\end{figure}


\end{document}


%
\maketitle

\section*{\large Characterization of the local oscillator and spectral purity transfer}

Figs. 5 and 6 depict characterization measurements of the frequency comb and local oscillator described in the Methods section.  Table 1 includes the parameter values for the  resonant features in the laser noise model given in Eqn. 3 of the Methods section.

\begin{figure}[!htb]
\centering
\includegraphics[width=1.0\textwidth]{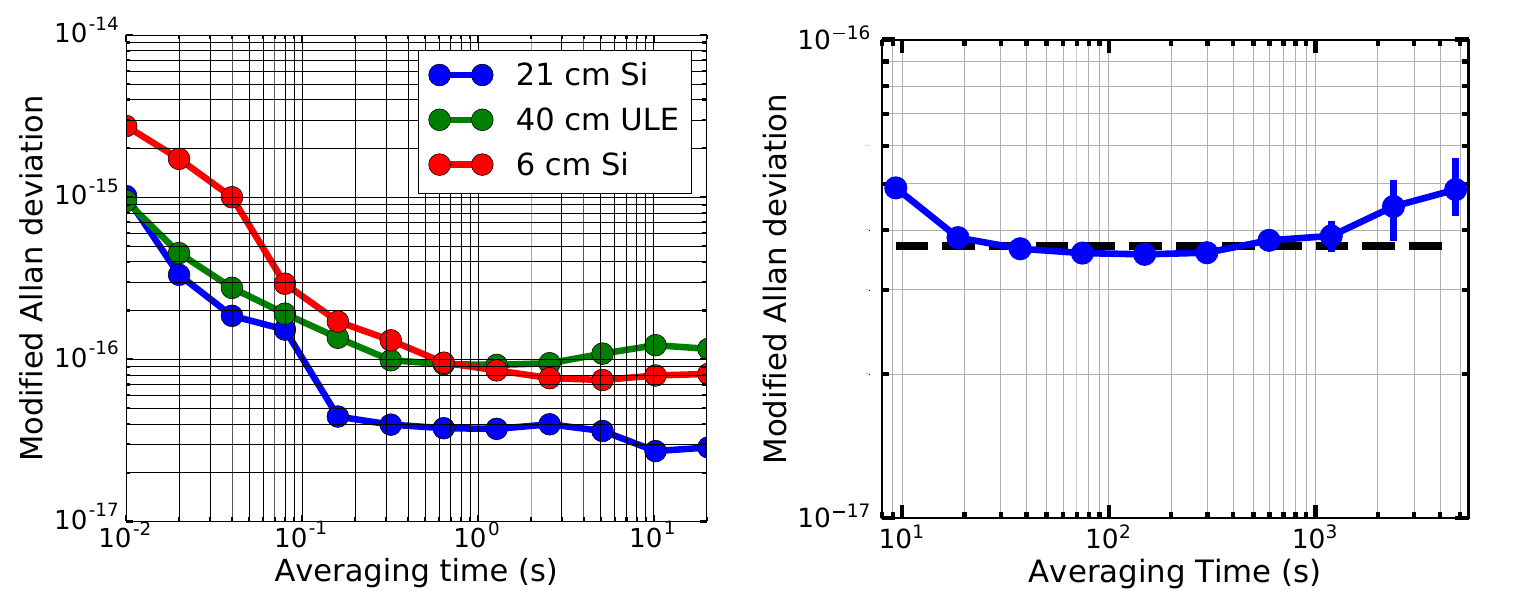}
{\caption*{ \label{fig:s}\textbf{Figure 5 |}  \textbf{Left:}  A three-cornered hat comparison between the local oscillator (21 cm Si) and two reference lasers based on a 6 cm Si cavity operating at 4K and a 40 cm cavity operating at room temperature characterizes the performance of the local oscillator at short averaging times (<10 s).  The data is taken with a zero dead-time lambda-type frequency counter operating with a 10-ms gate time.  \textbf{Right:}  The typical single-day stability of the local oscillator at longer averaging times is evaluated using the atomic servo correction signal for the 1D Sr lattice clock (30s servo attack time).  The stability is determined by averaging datasets from 13 days that are each at least 20000 seconds in length after fitting and removing a linear frequency drift from each dataset.  On average, the single-day datasets exibit thermal noise limited stability out to 1000 s, though one dataset showed no departure from thermal noise out to 10000 s (see Fig 2b).  The black dashed line is a flicker frequency noise floor of $3.7\times 10^{-17}$ obtained by fitting the Allan deviation between 40-1000 seconds.  This is in good agreement with the previously published value of $4\times 10^{-17}$\cite{Matei2017}.}}
\end{figure}

\begin{figure}[!htb]
\centering
\includegraphics[width=1.0\textwidth]{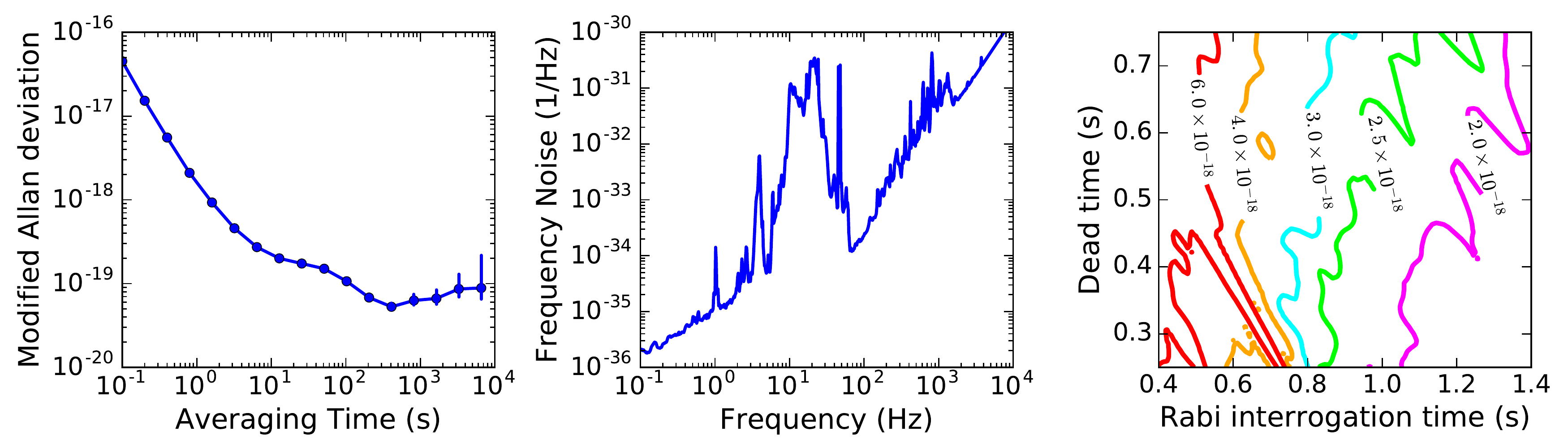}
{\caption*{ \label{fig:5}\textbf{Figure 6 |}  \textbf{Left:}  Modified Allan deviation of the estimated instability added to the local oscillator by the optical frequency comb.  This is inferred by performing a multi-line comparison measurement against an identical frequency comb\cite{Giunta2018} in order to characterize the differential noise between their 1542 nm and 698 nm outputs.  The two systems are stabilized to a common reference at 1542.14 nm and the hetrodyne beat between the systems at 698 nm is counted using a dead-time free Lambda-type frequency counter with a 100 ms gate time.  Only a few tens of centimeters of fiber are differential between the 1542 nm and 698 nm paths.  This allows both combs to achieve an unprecedented fractional instability of $1.6 \times 10^{-18}$ at one second.  The combs flicker at the high $10^{-20}$ level beyond 300 seconds due to thermally induced phase drifts occurring in uncompensated fiber paths.  \textbf{Center:} Phase noise power spectral density of the two-comb 698 nm hetrodyne beat in units of fractional frequency noise.  The accumulated optical phase drift observed over a seven hour measurement is only $1.5 \times 2\pi$, which corresponds to a fractional frequency offset with respect to the 698 nm carrier of $1.3\times10^{-19}$.  The large feature between 5-50 Hz arises due to acoustic noise present at the time of the qualification measurement.  Since relocating the system to JILA, the vibrational and acoustic isolation has been improved significantly. By analyzing the three-cornered hat comparison data from Fig. 5 in the frequency domain one infers that this spectral feature has been reduced by at least an order of magnitude since, if present, it would appear to limit the stability of the 40 cm ULE cavity at these Fourier frequencies.  \textbf{Right:}  The contribution from the comb to the Dick effect limit for clock stability is computed by inserting the measured frequency noise power spectral density into Eqn. 2 in the main text.  For the operating conditions used here (550-600 ms interrogation time and 570 ms dead time), the comb contribution is negligible.}}

\end{figure}
\newpage
\begin{singlespace}
\noindent {\normalsize \textbf{Table 1 |} Eqn. 3 in Methods presents the functional form of the local oscillator noise model.  The parameters corresponding to the resonant features in this model are tabulated below.}
\end{singlespace}

\begin{center}
 \begin{tabular}{||c c c c||} 
 \hline
 Index & $f_i$ (Hz) & $a_i$ (1/Hz) & $\Gamma_i$ (Hz) \\ [0.5ex] 
 \hline\hline
 1 & 5.7 & 7.0e-34 & 1.0 \\ 
 \hline
 2 & 12.7 & 1.5e-34 & 1.5 \\
 \hline
 3 & 20.0 & 4.0e-34 & 0.1 \\
 \hline
 4 & 30.0 & 5.0e-34 & 0.1 \\
 \hline
 5 & 40.0 & 5.0e-34 & 0.1 \\
 \hline
 6 & 45.0 & 1.0e-34 & 4 \\
 \hline
 7 & 55.0 & 4.0e-34 & 1.2 \\ [1ex] 
 \hline
\end{tabular}
\end{center}

\section*{\large Magnetic field noise rejection from a four-point locking sequence}

As discussed in the main text, fluctuations in the magnetic field background in the laboratory will induce Zeeman shifts in the clock transition due to the hyperfine structure of $^{87}\textrm{Sr}$.  Therefore a measurement of the detuning of the laser from resonance, $\Delta[n]$ where $n$ is the experimental cycle index, will include an additional contribution from the magnetic field background:

\begin{equation}
\Delta[n] = \Delta_{\textrm{laser}}[n] + \Delta_{\textrm{field}}[n]
\end{equation}

For the present discussion all other sources of technical noise are omitted for the sake of simplicity.  We make the further simplifying assumption that neither the laser nor the background field exibit long-term drift so we can repeatedly interrogate the clock transition without active feedback.  This allows one to neglect the impact of the servo on the frequency record of the clock.  The purpose of the present discussion is to derive the impact of short term noise from the laser and background field on the clocks frequency record.

\begin{figure}
\centering
\includegraphics[width=1.0\textwidth]{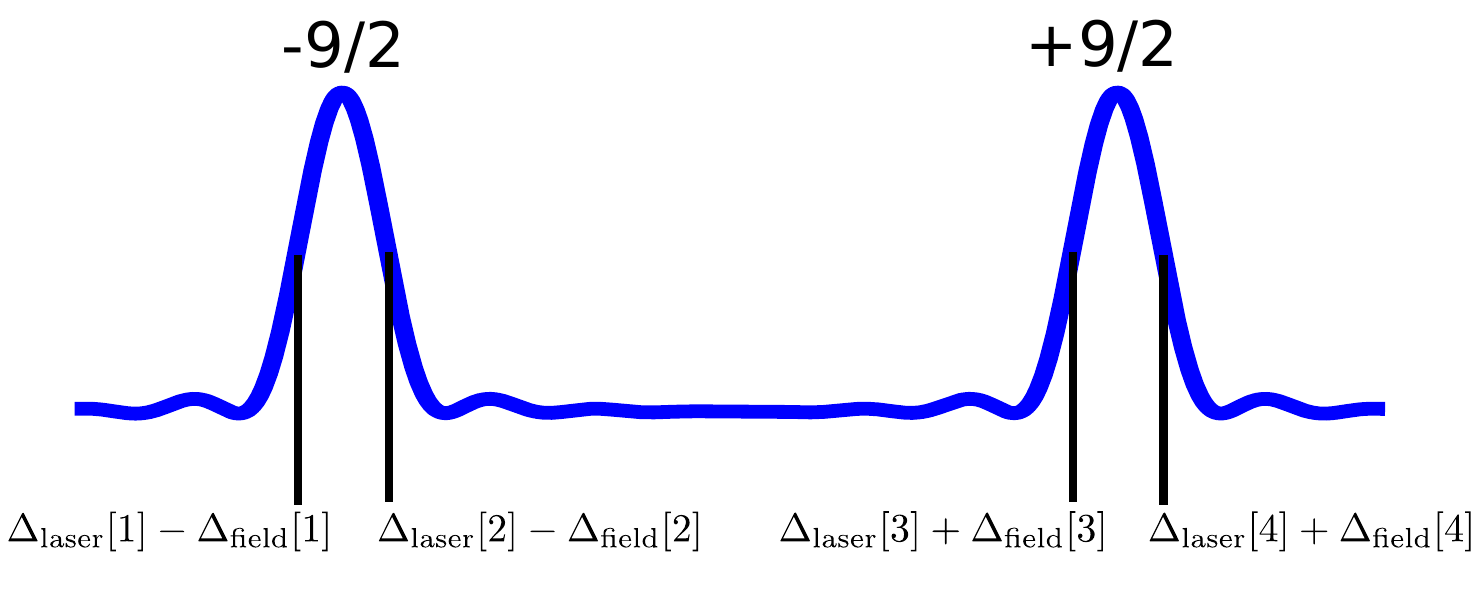}
{\caption*{ \label{fig:6}\textbf{Figure 7 |}  Diagram depicting the frequency measurements required for a four-point lock in the 1D lattice clock.  By averaging these four measurements one may reject frequency noise arising from first-order Zeeman shifts induced by slowly varying magnetic fields in the laboratory.}}
\end{figure}

To reject the magnetic field noise noise in the 1D clock, the detuning of the laser from resonance is measured using a four-point spectroscopy sequence as described in the main text. The four point measurement sequence for the 1D lattice clock is depicted in Fig. 7.  The important thing to note is that the first-order Zeeman shifts on the $\ket{^{1}S_{0},m_F=\pm \frac{9}{2}} \rightarrow \ket{^{3}P_{0},m_F=\pm \frac{9}{2}}$ transitions will be anticorrelated and should largely cancel when all four measurements are averaged provided that the magnetic field doesn't change dramatically over the course of the measurements (in this work, fluctuations in the second-order Zeeman shift are too small to have a significant impact on the clock stability and are ignored in the present discussion).  The discussion that follows also applies to the spectroscopy sequence used in the 3D clock, though in that case $\Delta_{\textrm{field}}$ is intrinsically a factor of 22 smaller \cite{Boyd2007}.  These four measurements yield the following

\begin{align}
\Delta[1] &= \Delta_{\textrm{laser}}[1] - \Delta_{\textrm{field}}[1] \\
\Delta[2] &= \Delta_{\textrm{laser}}[2] - \Delta_{\textrm{field}}[2] \\
\Delta[3] &= \Delta_{\textrm{laser}}[3] +\Delta_{\textrm{field}}[3]\\
\Delta[4] &= \Delta_{\textrm{laser}}[4] + \Delta_{\textrm{field}}[4] \\
\left\langle\Delta_{\textrm{laser}}\right\rangle &\approx \frac{1}{4}\left( \Delta[1] + \Delta[2] + \Delta[3] + \Delta[4] \right)
\end{align}

\noindent where $\left\langle\Delta_{\textrm{laser}}\right\rangle$ denotes the average laser detuning from resonance over the span of all four measurements and the sign convention chosen for $\Delta_{\textrm{field}}$ corresponds to the $\ket{^{1}S_{0},m_F= \frac{9}{2}} \rightarrow \ket{^{3}P_{0},m_F= \frac{9}{2}}$ transition.  The frequency record for the laser is then estimated by averaging successive four-point measurement cycles

\begin{equation}
\widehat{\Delta}_{\textrm{laser}} = \left\{... , \frac{1}{4}\left( \Delta[1] + \Delta[2] + \Delta[3] + \Delta[4] \right), \frac{1}{4}\left( \Delta[5] + \Delta[6] + \Delta[7] + \Delta[8] \right), ...\right\}
\end{equation}

\noindent In order to understand the impact of laser and field noise on $\widehat{\Delta}_{\textrm{laser}}$, one can derive transfer functions $H[\omega_{m}$] and $G[\omega_{m}$] corresponding to the coupling of laser and field noise respectively in the Fourier domain.  To this end, it is useful to point out that the sequence in Eqn. 7 can be constructed by applying the following two operations to the laser and field noise.

\noindent \textbf{Step 1:}  Filter the laser and field noise and add the resulting sequences:
\begin{equation}
\Delta_{\textrm{filt}}[n] = h*\Delta_{\textrm{laser}}[n] + g*\Delta_{\textrm{field}}[n]
\end{equation}

\noindent where $*$ denotes the convolution operator and

\begin{align}
h[n] &= \frac{1}{4}\sum_{k=0}^{3} \delta[n-k]\\
g[n] &= \frac{1}{4}\left(-\delta[n] - \delta[n-1] + \delta[n-2] + \delta[n-3]\right)
\end{align}

\noindent are discrete-time filter functions corresponding to the desired transfer functions. This operation yields

\begin{multline}
\left\{... ,\Delta[1],\Delta[2],\Delta[3],\Delta[4],\Delta[5],\Delta[6],...\right\} \\ \rightarrow \left\{...,\frac{1}{4}\left( \Delta[1] + \Delta[2] + \Delta[3] + \Delta[4] \right), \frac{1}{4}\left( \Delta[2] + \Delta[3] + \Delta[4] + \Delta[5] \right), ...\right\} \\
 = \left\{... ,\Delta_{\textrm{filt}}[4], \Delta_{\textrm{filt}}[5],...\right\}
\end{multline}

\noindent \textbf{Step 2.}  Downsample the data by a factor of four.  

\begin{multline}
\left\{... , \frac{1}{4}\left( \Delta[1] + \Delta[2] + \Delta[3] + \Delta[4] \right), \frac{1}{4}\left( \Delta[2] + \Delta[3] + \Delta[4] + \Delta[5] \right), ...\right\} \rightarrow \\
\left\{... , \frac{1}{4}\left( \Delta[1] + \Delta[2] + \Delta[3] + \Delta[4] \right), \frac{1}{4}\left( \Delta[5] + \Delta[6] + \Delta[7] + \Delta[8] \right), ...\right\} = \widehat{\Delta}_{\textrm{laser}}
\end{multline}

\noindent The impact of both magnetic field noise and laser noise on $\widehat{\Delta}_{\textrm{laser}}$ can be understood by considering the discrete Fourier transform of the frequency record $\Delta_{\textrm{filt}}$ followed by a reduction in bandwidth due to downsampling.  

\begin{equation}
\Delta_{\textrm{filt}}[\omega_m] = H[\omega_m]\Delta_{\textrm{laser}}[\omega_m] + G[\omega_m]\Delta_{\textrm{field}}[\omega_m]
\end{equation}

\noindent where $\omega_m = \frac{m \omega_s}{N_s}$, $\omega_s = 2\pi/T_c$ is our sampling rate in angular frequency units, $T_c$ is our experimental cycle time, and $N_s$ is the total number of samples, and $H[\omega_m]$ and $G[\omega_m]$ denote the discrete Fourier transforms of the filter functions $h[n]$ and $g[n]$ respectively.  

\begin{align}
H[\omega_m] &= \sum_{n=0}^{\infty} h[n]e^{-j\omega_m n} = \frac{1}{4}\sum_{k=0}^{3} e^{-j\omega_m k} =  \frac{1}{4} \frac{1-e^{-j4\omega_m}}{1-e^{-j\omega_m}} = \frac{1}{4} \frac{e^{-j2\omega_m}}{e^{-j\omega_m/2}} \frac{\sin\left(2\omega_m \right)}{\sin\left(\omega_m/2 \right)} \\
G[\omega_m] &= \frac{1}{4}\left(-1 -e^{-j\omega_m}+ e^{-j2\omega_m} + e^{-j3\omega_m} \right)
\end{align}

\begin{figure}
\centering
\includegraphics[width=1.0\textwidth]{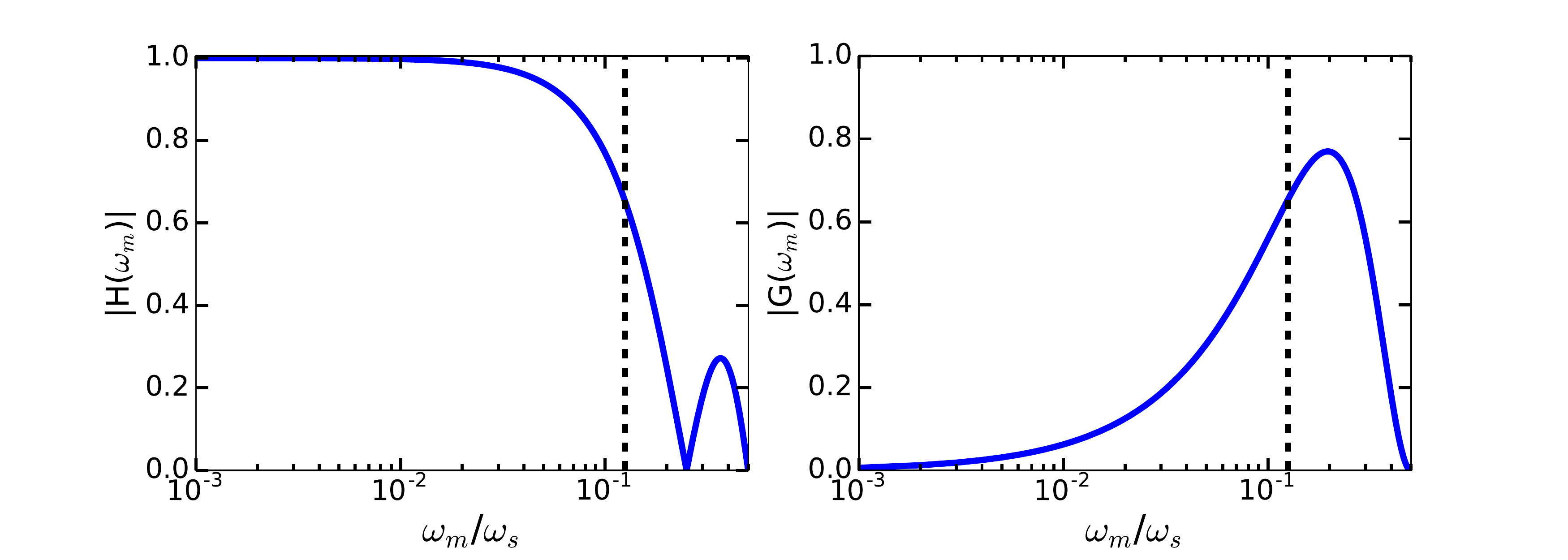}
{\caption*{ \label{fig:TFs}\textbf{Figure 8 | }Transfer functions corresponding to the coupling of laser noise (left) and magnetic field noise (right) to the frequency record of the clock when using the four-point interrogation sequence depicted in Fig. 7.  The x-axis is normalized by the clock cycle frequency $\omega_s$.  The dashed lines indicate the Nyquist frequency after performing the four-point analysis.} }
\end{figure}

$H[\omega_m]$ and $G[\omega_m]$ are simply transfer functions from laser and magnetic field noise to our frequency record.  The magnitudes of both filters are plotted in Fig. 8.  We see that $\Delta_{\textrm{laser}}[\omega_m]$ is effectively low-passed while $\Delta_{\textrm{field}}[\omega_m]$ is high passed.  Therefore, a four-point interrogation sequence will reject magnetic field noise provided that the fluctuations in the field are sufficiently slow.  After downsampling the data, the bandwidth of the coupled noise is reduced by a factor of four.  The dashed lines in Fig. 8 indicate the Nyquist frequency after downsampling.

Note that, at Fourier frequencies approaching the Nyquist rate, laser and magnetic field induced noise couple to the frequency record with almost equal magnitude.  This spectroscopy sequence is therefore not effective at rejecting high frequency magnetic field induced noise as it is unable to distinguish between the two noise sources.

Along with a reduction in bandwidth, the downsampling of the data by a factor of four leads to an additional aliasing process.  One might assume that this would lead to an increase in the Dick effect limit since laser noise at Fourier frequencies $\omega_m = \frac{\omega_s}{4}$ and $\omega_m = \frac{\omega_s}{2}$ will be aliased down to DC.  In fact, the Dick effect limit remains the same since $H[\omega_m] = 0$ at these frequencies.  $H[\omega_m]$ therefore functions as an anti-aliasing filter.  However, $G[\frac{\omega_s}{4}]$ is non-zero, so magnetic field noise at this frequency can alias down to DC and degrade the long-term stability of the clock.